\documentclass[aps,prb,manuscript]{revtex4}
\usepackage{color}
\usepackage{amsmath}
\usepackage{graphicx}
\usepackage{amssymb}
\begin{document}
\draft
\title{Analytic approach for calculating transmitted electromagnetic fields through a slot array in deep sub-wavelength regime}

\author{L. David Wellems and Danhong Huang}

\address{Air Force Research Laboratory, Space Vehicles
Directorate,\\
Kirtland Air Force Base, NM 87117, USA}

\date{\today}

\begin{abstract}
For the diffraction of an incident plane electromagnetic wave by a slotted metallic film,
the previous analytical calculation for a single slot [Technical Phys. {\bf 50}, 1076 (2005)] is generalized into a model for an arbitrary linear array of slots with variable slot width, slot separation and
slot dielectric material.
The advantage as well as the effectiveness of the generalized model presented in this paper are best described by enabling calculation of a continuous spatial distribution of an electromagnetic field by inverting a small
discrete coefficient matrix spanned by both the slot index and the slot-eigenmode index for a set of linear equations.
In comparisons with well-known plane-wave and finite-difference time-domain methods,
inverting a large matrix, in wave number space for the former case and in real space at each time step for the latter case, can be avoided to greatly speed up numerical calculations.
In addition, based on a partial-domain method, the formalism presented here can be employed to treat a composite surface (e.g., a slotted metal film with different dielectric materials in the slots),
while the analytical Green's function approach [J. Opt. A: Pure Appl. Opt. {\bf 8}, S191 (2006)] becomes intractable in this case. Some numerical results are presented as a demonstration
of this new analytical model.
\end{abstract}


\maketitle

\section{Introduction}
\label{sec1}

For a periodically-patterned surface on a metal film, the Maxwell equations can be solved by using a plane-wave (PW) method,\,\cite{wendler,huang,huang1} or a finite-difference time-domain (FDTD) method,\,\cite{yee,fan,fdtd} or a Green's function (GF) method.\,\cite{maradudin,maradudin1} For the PW method, a large matrix in wave number space with respect to different reciprocal lattice vectors needs to be inverted in order to obtain
the spatial distribution of the total electromagnetic (EM) field. For the FDTD method, the time evolution into a steady state EM field is obtained through inverting a large matrix in real space at each time
step with respect to all grid points in a region containing a unit-cell and a surrounding space not far from the surface. For the GF method, all the values of the EM field on a smooth surface must be calculated
by inverting a large matrix in real space with respect to all grid points on the boundaries to get the spatial distribution of the total EM field. Both the PW and GF methods have an advantage for calculating a far-field
distribution, while the FDTD method is usually limited to the calculation of the near-field distribution. Moreover, the GF method can also be used for a non-periodic smooth surface profile function.\,\cite{maradudin}
\medskip

For a non-smooth surface, on the other hand,
such as a one-dimensional periodic slot array, the GF method cannot be used. In this case, however, the partial-domain method\,\cite{maradudin,Mittra} used in the GF theory can still be employed
in combination with the slot-eigenmode expansion\,\cite{wendler} as well as with
the Fourier expansion\,\cite{huang,huang1} methods. Based on this technique, an analytic solution for the diffraction of a plane EM wave by a single slot
has been obtained.\,\cite{Serdyuk} This analytical single-slot approach has been generalized in this paper to deal with an arbitrary slot array.
If a single slot is replaced by double slots, each surface-plasmon-polariton branch\,\cite{ebbesen,ebbesen1,ebbesen2} will be split into two with a minigap controlled by an EM coupling between the two slots.\,\cite{Wellems}
In the case of double slots with different widths and filled with different dielectric materials, the circulation and weaving of light was found as a result of the excitation of a phase resonance.\,\cite{crouse}
\medskip

In this paper, a two-dimensional (2D) theoretical model for the diffraction of a plane EM wave by a perfectly electrically  conducting (PEC) slotted film is derived
with variable slot width, slot separation and slot dielectric
material. The solutions of the 2D Helmholtz equation in different partial domains are matched to each other by proper boundary conditions for both PEC and slot dielectric medium.
This method has applications in modeling surface plasmon polaritrons or surface waves,\,\cite{hou}
optical beam steering,\,\cite{lee} and metamaterial design.\,\cite{gvidal}
The use of a PEC material precludes modeling of dissipation but allows for simple modal expressions for the slot fields.\,\cite{wendler,Serdyuk}
When the optical depth of a metallic film is small and energy dissipation is not a consideration (off-resonance),
the use of a PEC partial domain model is expected to be a good approximation.
\medskip

This paper is organized as follows. In Sec.\,\ref{sec2}, we generalize the single-slot model into one for an arbitrary array of slots with both $p$ and $s$ polarizations.
In Sec.\,\ref{sec3}, numerical results as a demonstration are presented and compared for the transmitted EM wave with $p$ polarization through a slot array with different slot numbers and incident angles.
A brief summary of the generalized model is given in Sec.\,\ref{sec4}.

\section{Theory}
\label{sec2}

In this paper, the incident EM field is assumed to be a plane wave with a simple time dependence proportional to $e^{-i\omega t}$, where $\omega$ is the angular frequency of the incident field.
For the $s$-polarized case, we write ${\bf E}=(0,\,E_y,\,0)\equiv (0,\,u,\,0)$ for the transverse electric field and ${\bf H}=[-i/(\omega\mu_0)]\,\overrightarrow{\nabla}\times{\bf E}$.
For the $p$-polarized case, on the other hand, we write ${\bf H}=(0,\,H_y,\,0)\equiv (0,\,u,\,0)$ and ${\bf E}=[i/(\omega\epsilon_0\epsilon)]\,\overrightarrow{\nabla}\times{\bf H}$.
Here, $\epsilon_0$ and $\mu_0$ are the vacuum permittivity and permeability,
$c=1/\sqrt{\epsilon_0\,\mu_0}$ is the speed of light in vacuum, and $\epsilon$ is the relative dielectric constant of a non-magnetic dielectric medium.
\medskip

The master equation for determining the scalar function $u$ introduced above is the two-dimensional Helmholtz equation in the dielectric medium in the $xz-$plane:

\begin{equation}
\frac{\partial^2 u}{\partial x^2} + \frac{\partial^2 u}{\partial z^2} + k^2\,u=0\ ,
\label{helm}
\end{equation}
where $k=(\omega/c)\sqrt{\epsilon}$ is the wave number of the field.
For a perfect electric conductor (PEC) with an extremely-large conductivity,
the boundary condition for $s$ polarization associated with Eq.\,(\ref{helm}) is just $u=0$ along all PEC boundaries. In addition,
for $p$ polarization, the boundary condition of Eq.\,(\ref{helm}) is $ \partial u/ \partial n=0$ along all PEC boundaries
where $n$ is in the surface normal direction.
\medskip

\subsection{$s-$Polarization}
\label{sec2.1}

\begin{figure}[p]
\begin{center}
\includegraphics*[width=6.0in]{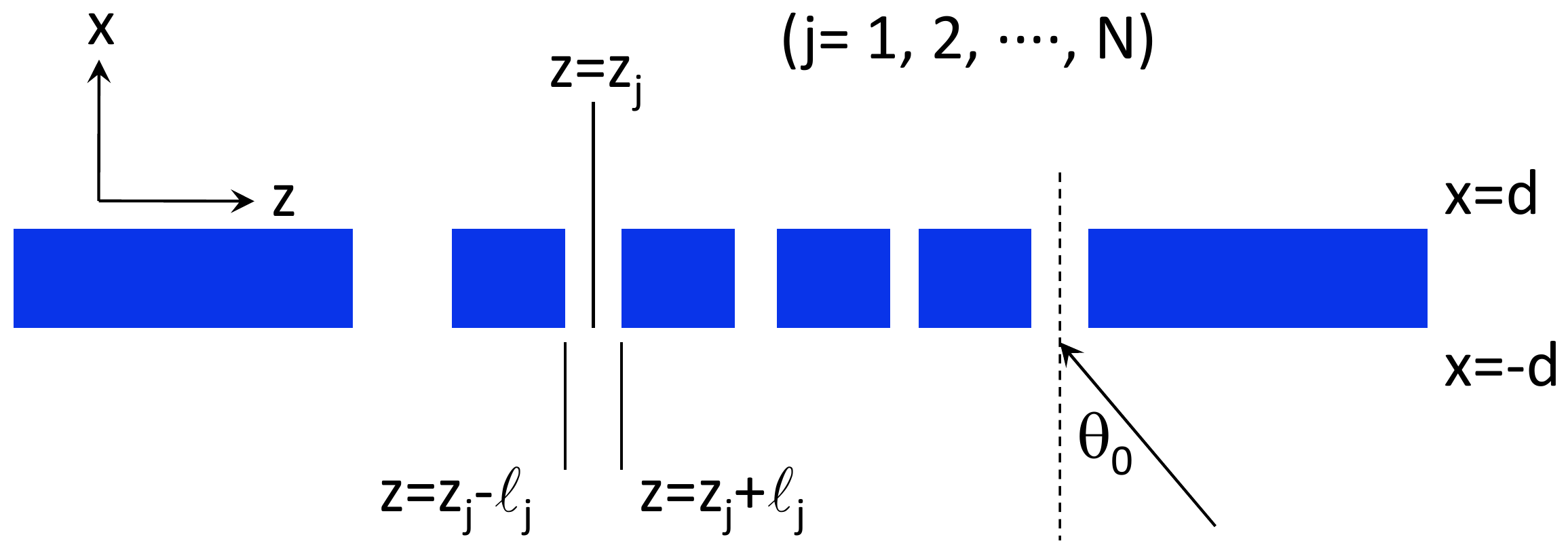}
\caption{(Color online) Illustration of a one-dimensional slot array in the $xz-$plane, where $j=1,\,2,\,\cdots,\,N$ is the slot index,
$z_j$ is the center of $j$th slot, $2\ell_j$ is the $j$th slot width, the slot depth is $2d$, and $\theta_0$ is the incident angle. The regions below and above the slot array are defined as Region I and Region III, respectively.
The slot-array region is designated Region II in the text.}
\label{f1}
\end{center}
\end{figure}

For the system shown in Fig.\,\ref{f1}, Region I, below the slotted layer ($x<-d$), has the boundary condition

\begin{equation}
\left.u\right|_{x=-d-0}=\left\{
\begin{array}{ll}
0\ , &\ \ \ \ \mbox{for upper PEC\ \ \ \ $z_j+\ell_j<z<z_{j+1}-\ell_{j+1}$}\\
\left.u\right|_{x=-d+0}\ ,&\ \ \ \ \mbox{for middle slot\ \ \ \ $|z-z_j|<\ell_j$}\\
0\ , &\ \ \ \ \mbox{for lower PEC\ \ \ \ $z_{j-1}+\ell_{j-1}<z<z_j-\ell_j$}
\end{array}\right.\ ,
\label{sboundary-1}
\end{equation}
where $j=1,\,2,\,\cdots,\,N$ is the slot index, $z_j$ is the coordinate of the $j$th slot center, $2\ell_j$ is the
width of the $j$th slot, and $2d$ is the depth of all slots or screen thickness.
For $j=N$, we take $z_{N+1}=\infty$ and $\ell_{N+1}=0$ in Eq.\,(\ref{sboundary-1}). Similarly, $z_{0}=-\infty$ and $\ell_{0}=0$
in Eq.\,(\ref{sboundary-1}) when $j=1$.
Moreover, from the continuity of the derivative of $u$, we have another boundary condition

\begin{equation}
\,\left.\frac{\partial u}{\partial x}\right|_{x=-d-0}=
\,\left.\frac{\partial u}{\partial x}\right|_{x=-d+0}\ ,\ \ \ \ \mbox{for all middle slots\ \ \ \ $|z-z_j|<\ell_j$}\ .
\label{sboundaryd-1}
\end{equation}
In a similar way, we are able to get the boundary conditions for Region III above the slotted layer ($x>d$) from Eqs.\,(\ref{sboundary-1}) and (\ref{sboundaryd-1})
by setting $x= d\pm 0$.
Finally, the field must be zero along the interior slot walls in Region II ($|x|\leq d$), that is,

\begin{equation}
u(x,\,z=z_j\pm\ell_j)=0\ ,\ \ \ \ \mbox{at all slot walls\ \ \ \ $|x|\leq d$}\ .
\label{e4}
\end{equation}
\medskip

In Region I (below the slot layer), the solution of the Helmholtz equation in Eq.\,(\ref{helm}), including the incident field (incident from below), can be written as

\begin{eqnarray}
u^{({\rm I})}(x,\,z)&=&e^{ik_0(x+d)}\,e^{i\beta_0z}-e^{-ik_0(x +d)}\,e^{i\beta_0z}
\nonumber\\
&+&\int\limits_0^{\infty} d\beta\,\left[A_s(\beta)\,\cos(\beta z)+
iA_a(\beta)\,\sin(\beta z )\right]\,e^{-ik_1(\beta)(x + d)}\ ,
\label{sFourier1}
\end{eqnarray}
where the subscripts $s$ and $a$ represent the symmetric and antisymmetric contributions, separately,
$k_0=(\omega n_{\rm L}/c)\,\cos\theta_0$, $\beta_0=(\omega n_{\rm L}/c)\,\sin\theta_0$,
$\beta$ is real, $k_1(\beta)=\sqrt{(\omega n_{\rm L}/c)^2-\beta^2}$ can be either real or complex with ${\sf Im}[k_1(\beta)]\geq 0$,
$n_{\rm L}=\sqrt{\epsilon_{\rm L}}$ is the refractive index for the dielectric medium below the slot layer,
and $\theta_0$ is the incident angle. In Region III (above the slot layer), on the other hand,
the solution of the Helmholtz equation takes the form of

\begin{equation}
u^{({\rm III})}(x,\,z)=\int\limits_0^{\infty} d\beta\,\left[B_s(\beta)\,\cos(\beta z) +
iB_a(\beta)\,\sin(\beta z)\right]\,e^{ik_2(\beta)(x - d)}\ ,
\label{sFourierR}
\end{equation}
where $n_{\rm R}=\sqrt{\epsilon_{\rm R}}$ is the refractive index for the dielectric medium above the slots and
$k_2(\beta)=\sqrt{(\omega n_{\rm R}/c)^2-\beta^2}$ can also be real or complex with ${\sf Im}[k_2(\beta)]\geq 0$.
\medskip

In Region II (the slot layer), the solution of the Helmholtz equation along with the boundary condition in Eq.\,(\ref{e4}) is found to be

\begin{eqnarray}
u^{({\rm II})}(x,\,z)&=&\sum_j\,\theta(\ell_j-|z-z_j|)\sum_n\,\left\{\left[a_{sn}^j\,e^{i\sigma_{sn}^j(d+x)}+b_{sn}^j\,e^{i\sigma_{sn}^j(d-x)}\right]\,\cos[\xi_{sn}^j(z-z_j)]\right.
\nonumber\\
&+&\left.i\left[a_{an}^j\,e^{i\sigma_{an}^j(d+x)}+ b_{an}^j\,e^{i\sigma_{an}^j(d-x)}\right]\,\sin[\xi_{an}^j(z-z_j)]\right\}\ ,
\label{smode1}
\end{eqnarray}
where $\theta(z)$ is the unit step function, $n=1,\,2,\,\cdots$ is an integer for the eigenmode index, $\xi_{sn}^j=(\pi/\ell_j)\,(n-1/2)$ and $\xi_{an}^j=(n\pi/\ell_j)$
are for the symmetric and antisymmetric slot modes, respectively,
$\sigma^j_{sn,\,an}=\sqrt{(\omega n_j/c)^2-(\xi_{sn,\,an}^j)^2}$ can be either real or complex with ${\sf Re}[\sigma^j_{sn,\,an}]\geq 0$, and $n_j=\sqrt{\epsilon_j}$ is the refractive index for the
dielectric medium inside the $j$th slot.
\medskip

By using the boundary conditions in Eq.\,(\ref{sboundary-1}) for $u$ at $x=\pm\,d$ and the orthogonality of the Fourier expansions in Eqs.\,(\ref{sFourier1}) and (\ref{sFourierR}),
the coefficients for both symmetric and antisymmetric contributions can be expressed by the slot eigenmodes as follows:

\begin{eqnarray}
A_s(\beta)&=&\frac{1}{\pi}\int\limits_{-\infty}^{\infty} dz\,u^{({\rm II})}(x=-d,\,z)\,\cos(\beta z)
\nonumber\\
&=&\sum_n\,\left\{\sum_j\,\frac{\ell_j}{\pi}\,\left[\left(a_{sn}^j+b_{sn}^j\,e^{2i\sigma_{sn}^jd}\right)\,Q_{sn}^j(\beta)\,\cos(\beta z_j)\right.\right.
\nonumber\\
&-&\left.\left.i\left(a_{an}^j+ b_{an}^j\,e^{2i\sigma_{an}^jd}\right)\,Q_{an}^j(\beta)\,\sin(\beta z_j)\right]\right\}\ ,
\label{CoefAs}
\end{eqnarray}

\begin{eqnarray}
iA_a(\beta)&=&\frac{1}{\pi}\int\limits_{-\infty}^{\infty} dz\,u^{({\rm II})}(x=-d,\,z)\,\sin(\beta z)
\nonumber\\
&=&\sum_n\,\left\{\sum_j\,\frac{\ell_j}{\pi}\,\left[\left(a_{sn}^j+b_{sn}^j\,e^{2i\sigma_{sn}^jd}\right)\,Q_{sn}^j(\beta)\,\sin(\beta z_j)\right.\right.
\nonumber\\
&+&\left.\left.i\left(a_{an}^j+b_{an}^j\,e^{2i\sigma_{an}^jd}\right)\,Q_{an}^j(\beta)\,\cos(\beta z_j)\right]\right\}\ ,
\label{CoefAa}
\end{eqnarray}

\begin{eqnarray}
B_s(\beta)&=&\frac{1}{\pi}\int\limits_{-\infty}^{\infty} dz\,u^{({\rm II})}(x=d,\,z)\,\cos(\beta z)
\nonumber\\
&=&\sum_n\,\left\{\sum_j\,\frac{\ell_j}{\pi}\,\left[\left(a_{sn}^j\,e^{2i\sigma_{sn}^jd}+b_{sn}^j\right)\,Q_{sn}^j(\beta)\,\cos(\beta z_j)\right.\right.
\nonumber\\
&-&\left.\left.i\left(a_{an}^j\,e^{2i\sigma_{an}^jd}+b_{an}^j\right)\,Q_{an}^j(\beta)\,\sin(\beta z_j)\right]\right\}\ ,
\label{CoefBs}
\end{eqnarray}

\begin{eqnarray}
iB_a(\beta)&=&\frac{1}{\pi}\int\limits_{-\infty}^{\infty} dz\,u^{({\rm II})}(x=d,\,z)\,\sin(\beta z)
\nonumber\\
&=&\sum_n\,\left\{\sum_j\,\frac{\ell_j}{\pi}\,\left[\left(a_{sn}^j\,e^{2i\sigma_{sn}^jd}+b_{sn}^j\right)\,Q_{sn}^j(\beta)\,\sin(\beta z_j)\right.\right.
\nonumber\\
&+&\left.\left.i\left(a_{an}^j\,e^{2i\sigma_{an}^jd}+b_{an}^j\right)\,Q_{an}^j(\beta)\,\cos(\beta z_j)\right]\right\}\ .
\label{CoefBa}
\end{eqnarray}
In Eqs.\,(\ref{CoefAs})-(\ref{CoefBa}), we have defined the notations

\begin{eqnarray}
Q_{sn}^j(\beta)&=&
\frac{1}{\ell_j}\int\limits_{-\ell_j}^{\ell_j} dz\,\cos(\beta z)\,\cos(\xi_{sn}^jz)
\nonumber\\
&=&{\rm sinc}[(\beta-\xi_{sn}^j)\,\ell_j]+{\rm sinc}[(\beta+\xi_{sn}^j)\,\ell_j]\ ,
\label{Qss}
\end{eqnarray}

\begin{eqnarray}
Q_{an}^j(\beta)&=&
\frac{1}{\ell_j}\int\limits_{-\ell_j}^{\ell_j} dz\,\sin(\beta z)\,\sin(\xi_{an}^jz)
\nonumber\\
&=&{\rm sinc}[(\beta-\xi_{an}^j)\,\ell_j]-{\rm sinc}[(\beta+\xi_{an}^j)\,\ell_j]\ ,
\label{Qaa}
\end{eqnarray}
where ${\rm sinc}(x)\equiv\sin(x)/x$.
\medskip

The slot eigenmode expansion coefficients $a_{sn}^j$, $b_{sn}^j$, $a_{an}^j$ and $b_{an}^j$ in Eq.\,(\ref{smode1}) are determined from the derivative
continuities in Eq.\,(\ref{sboundaryd-1}) for $u$ at $x=\pm\,d$ and for each slot.
Therefore, we get for $j=1,\,2,\,\cdots,\,N$

\begin{eqnarray}
&&\left.\int\limits_{z_j-\ell_j}^{z_j+\ell_j} \frac{dz}{\ell_j}\,
\left[\begin{array}{c}
\cos[\xi_{sn}^j(z-z_j)]\\
\sin[\xi_{an}^j(z-z_j)]
\end{array}\right]\,
\frac{\partial u^{{\rm I}}(x,\,z)}{\partial x}\right|_{x=-d-0}
\nonumber
\\
\nonumber\\
&=&\left.\int\limits_{z_j-\ell_j}^{z_j+\ell_j} \frac{dz}{\ell_j}\,
\left[\begin{array}{c}
\cos[\xi_{sn}^j(z-z_j)]\\
\sin[\xi_{an}^j(z-z_j)]
\end{array}\right]\,
\frac{\partial u^{{\rm II}}(x,\,z)}{\partial x}\right|_{x=-d+0}\ ,
\label{boundarydd-2}
\end{eqnarray}

\begin{eqnarray}
&&\left.\int\limits_{z_j-\ell_j}^{z_j+\ell_j} \frac{dz}{\ell_j}\,
\left[\begin{array}{c}
\cos[\xi_{sn}^j(z-z_j)]\\
\sin[\xi_{an}^j(z-z_j)]
\end{array}\right]\,
\frac{\partial u^{{\rm III}}(x,\,z)}{\partial x}\right|_{x=d+0}
\nonumber
\\
\nonumber\\
&=&\left.\int\limits_{z_j-\ell_j}^{z_j+\ell_j} \frac{dz}{\ell_j}\,
\left[\begin{array}{c}
\cos[\xi_{sn}^j(z-z_j)]\\
\sin[\xi_{an}^j(z-z_j)]
\end{array}\right]\,
\frac{\partial u^{{\rm II}}(x,\,z)}{\partial x}\right|_{x=d-0}\ .
\label{boundarydd-3}
\end{eqnarray}
Using the orthogonality of the slot eigenmodes, from Eqs.\,(\ref{boundarydd-2}) and (\ref{boundarydd-3}) we arrive at the following set of equations:

\begin{eqnarray}
2k_0\,e^{i\beta_0z_j}\,Q^j_{sn}(\beta_0)&-&\int\limits_0^{\infty} d\beta\,\left[A_s(\beta)\,\cos(\beta z_j)+iA_a(\beta)\,
\sin(\beta z_j)\right]\,Q^j_{sn}(\beta)\,k_1(\beta)
\nonumber\\
&=&\sigma^j_{sn}\,(a^j_{sn}-b^j_{sn}\,e^{2i\sigma^j_{sn}d})\ ,
\label{deq1}
\end{eqnarray}

\begin{eqnarray}
-2k_0\,e^{i\beta_0z_j}\,Q^j_{an}(\beta_0)&+&\int\limits_0^{\infty} d\beta\,\left[iA_s(\beta)\,\sin(\beta z_j)+A_a(\beta)\,
\cos(\beta z_j)\right]\,Q^j_{an}(\beta)\,k_1(\beta)
\nonumber\\
&=&-\sigma^j_{an}\,(a^j_{an}-b^j_{an}\,e^{2i\sigma^j_{an}d})\ ,
\label{deq2}
\end{eqnarray}

\begin{equation}
\int\limits_0^{\infty} d\beta\,\left[B_s(\beta)\,\cos(\beta z_j)+iB_a(\beta)\,
\sin(\beta z_j)\right]\,Q^j_{sn}(\beta)\,k_2(\beta)
=\sigma^j_{sn}\,(a^j_{sn}\,e^{2i\sigma^j_{sn}d}-b^j_{sn})\ ,
\label{deq3}
\end{equation}

\begin{equation}
\int\limits_0^{\infty} d\beta\,\left[iB_s(\beta)\,\sin(\beta z_j)+B_a(\beta)\,
\cos(\beta z_j)\right]\,Q^j_{an}(\beta)\,k_2(\beta)
=\sigma^j_{an}\,(a^j_{an}\,e^{2i\sigma^j_{an}d}-b^j_{an})\ ,
\label{deq4}
\end{equation}
where $j=1,\,2,\,\cdots,\,N$.
\medskip

Inserting Eqs.\,(\ref{CoefAs})-(\ref{CoefBa}) into Eqs.\,(\ref{deq1})-(\ref{deq4}) gives rise to a set of $4N$ linear equations with respect to
$a_{sn}^j$, $b_{sn}^j$, $a_{an}^j$ and $b_{an}^j$. That is, for $j=1,\,2,\,\cdots,\,N$,

\begin{eqnarray}
&&\sum_{n^\prime,\,j^{\,\prime}}\,\left(a_{sn^\prime}^{j^{\,\prime}}+b_{sn^\prime}^{j^{\,\prime}}\,e^{2i\sigma_{sn^\prime}^{j^{\,\prime}}d}\right)\,
\left\{\frac{\ell_{j^{\,\prime}}}{\pi}\int\limits_0^{\infty} d\beta\,Q_{sn}^j(\beta)\,Q_{sn^\prime}^{j^{\,\prime}}(\beta)\,k_1(\beta)\,
\cos[\beta(z_{j^{\,\prime}}-z_j)]\right\}
\nonumber\\
&-&i\sum_{n^\prime,\,j^{\,\prime}}\,\left(a_{an^\prime}^{j^{\,\prime}}+b_{an^\prime}^{j^{\,\prime}}\,e^{2i\sigma_{an^\prime}^{j^{\,\prime}}d}\right)\,
\left\{\frac{\ell_{j^{\,\prime}}}{\pi}\int\limits_0^{\infty} d\beta\,Q_{sn}^j(\beta)\,Q_{an^\prime}^{j^{\,\prime}}(\beta)\,k_1(\beta)\,
\sin[\beta(z_{j^{\,\prime}}-z_j)]\right\}
\nonumber\\
&+&\sum_{n^\prime,\,j^{\,\prime}}\delta_{j,j^\prime}\delta_{n,n^\prime}\sigma^{j^\prime}_{sn^\prime}\,(a^{j^\prime}_{sn^\prime}-b^{j^\prime}_{sn^\prime}\,
e^{2i\sigma^{j^\prime}_{sn^\prime}d})=2k_0\,e^{i\beta_0z_j}\,Q^j_{sn}(\beta_0)\ ,
\label{seq5}
\end{eqnarray}

\begin{eqnarray}
&&-i\sum_{n^\prime,\,j^{\,\prime}}\,\left(a_{sn^\prime}^{j^{\,\prime}}+b_{sn^\prime}^{j^{\,\prime}}\,e^{2i\sigma_{sn^\prime}^{j^{\,\prime}}d}\right)\,
\left\{\frac{\ell_{j^{\,\prime}}}{\pi}\int\limits_0^{\infty} d\beta\,Q_{an}^j(\beta)\,Q_{sn^\prime}^{j^{\,\prime}}(\beta)\,k_1(\beta)\,
\sin[\beta(z_j-z_{j^{\,\prime}})]\right\}
\nonumber\\
&+&\sum_{n^\prime,\,j^{\,\prime}}\,\left(a_{an^\prime}^{j^{\,\prime}}+b_{an^\prime}^{j^{\,\prime}}\,e^{2i\sigma_{an^\prime}^{j^{\,\prime}}d}\right)\,
\left\{\frac{\ell_{j^{\,\prime}}}{\pi}\int\limits_0^{\infty} d\beta\,Q_{an}^j(\beta)\,Q_{an^\prime}^{j^{\,\prime}}(\beta)\,k_1(\beta)\,
\cos[\beta(z_j-z_{j^{\,\prime}})]\right\}
\nonumber\\
&+&\sum_{n^\prime,\,j^{\,\prime}}\delta_{j,j^\prime}\delta_{n,n^\prime}\sigma^{j^\prime}_{an^\prime}\,(a^{j^\prime}_{an^\prime}-b^{j^\prime}_{an^\prime}\,
e^{2i\sigma^{j^\prime}_{an^\prime}d})=2k_0\,e^{i\beta_0z_j}\,Q^j_{an}(\beta_0)\ ,
\label{seq6}
\end{eqnarray}

\begin{eqnarray}
&&\sum_{n^\prime,\,j^{\,\prime}}\,\left(a_{sn^\prime}^{j^{\,\prime}}\,e^{2i\sigma_{sn^\prime}^{j^{\,\prime}}d}+b_{sn^\prime}^{j^{\,\prime}}\right)\,
\left\{\frac{\ell_{j^{\,\prime}}}{\pi}\int\limits_0^{\infty} d\beta\,Q_{sn}^j(\beta)\,Q_{sn^\prime}^{j^{\,\prime}}(\beta)\,k_2(\beta)\,
\cos[\beta(z_{j^{\,\prime}}-z_j)]\right\}
\nonumber\\
&-&i\sum_{n^\prime,\,j^{\,\prime}}\,\left(a_{an^\prime}^{j^{\,\prime}}\,e^{2i\sigma_{an^\prime}^{j^{\,\prime}}d}+b_{an^\prime}^{j^{\,\prime}}\right)\,
\left\{\frac{\ell_{j^{\,\prime}}}{\pi}\int\limits_0^{\infty} d\beta\,Q_{sn}^j(\beta)\,Q_{an^\prime}^{j^{\,\prime}}(\beta)\,k_2(\beta)\,
\sin[\beta(z_{j^{\,\prime}}-z_j)]\right\}
\nonumber\\
&-&\sum_{n^\prime,\,j^{\,\prime}}\delta_{j,j^\prime}\delta_{n,n^\prime}\sigma^{j^\prime}_{sn^\prime}\,(a^{j^\prime}_{sn^\prime}\,e^{2i\sigma^{j^\prime}_{sn^\prime}d}-b^{j^\prime}_{sn^\prime})=0\ ,
\label{seq7}
\end{eqnarray}

\begin{eqnarray}
&&-i\sum_{n^\prime,\,j^{\,\prime}}\,\left(a_{sn^\prime}^{j^{\,\prime}}\,e^{2i\sigma_{sn^\prime}^{j^{\,\prime}}d}+b_{sn^\prime}^{j^{\,\prime}}\right)\,
\left\{\frac{\ell_{j^{\,\prime}}}{\pi}\int\limits_0^{\infty} d\beta\,Q_{an}^j(\beta)\,Q_{sn^\prime}^{j^{\,\prime}}(\beta)\,k_2(\beta)\,
\sin[\beta(z_j-z_{j^{\,\prime}})]\right\}
\nonumber\\
&+&\sum_{n^\prime,\,j^{\,\prime}}\,\left(a_{an^\prime}^{j^{\,\prime}}\,e^{2i\sigma_{an^\prime}^{j^{\,\prime}}d}+b_{an^\prime}^{j^{\,\prime}}\right)\,
\left\{\frac{\ell_{j^{\,\prime}}}{\pi}\int\limits_0^{\infty} d\beta\,Q_{an}^j(\beta)\,Q_{an^\prime}^{j^{\,\prime}}(\beta)\,k_2(\beta)\,
\cos[\beta(z_j-z_{j^{\,\prime}})]\right\}
\nonumber\\
&-&\sum_{n^\prime,\,j^{\,\prime}}\delta_{j,j^\prime}\delta_{n,n^\prime}\sigma^{j^\prime}_{an^\prime}\,(a^{j^\prime}_{an^\prime}\,e^{2i\sigma^{j^\prime}_{an^\prime}d}-b^{j^\prime}_{an^\prime})=0\ .
\label{seq8}
\end{eqnarray}
If we truncate the number of slot eigenmodes in the expansion, i.e., $n=1,\,2,\,\cdots,\,M$, this constitutes a $4NM$-order inhomogeneous linear-matrix equation, which can be exactly solved by inverting its
coefficient matrix.
Here the symmetric and antisymmetric eigenmodes in different slots with $z_j\neq z_{j^{\,\prime}}$ are coupled to each other, which is different from the single-slot result.\,\cite{Serdyuk}

\subsection{$p-$Polarization}
\label{sec2.2}

Following an approach that parallels the approach we performed above to deal with the boundary conditions for $s$ polarization, we require for $p$ polarization that the derivative condition is interchanged with the null
conditions, i.e., $u=0$ becomes  $\partial u/\partial n=0$ along the PEC surfaces. In addition,
the $(1/\epsilon)\,\partial u/\partial x$ continuity condition across the slot interfaces is used to determine the Fourier coefficients outside Region II
instead of the $u$ continuity condition. For Region I ($x<-d$), we have the boundary condition

\begin{equation}
\left. \frac{1}{\epsilon_{\rm L}}\frac{\partial u}{\partial x}\right|_{x=-d-0}=\left\{
\begin{array}{ll}
0\ , &\ \ \ \ \mbox{for upper PEC\ \ \ \ $z_j+\ell_j<z<z_{j+1}-\ell_{j+1}$}\\
\left.\frac{1}{\epsilon_j}\frac{\partial u}{\partial x}\right|_{x=-d+0}\ ,&\ \ \ \ \mbox{for middle slot\ \ \ \ $|z-z_j|<\ell_j$}\\
0\ , &\ \ \ \ \mbox{for lower PEC\ \ \ \ $z_{j-1}+\ell_{j-1}<z<z_j-\ell_j$}
\end{array}\right.\ .
\label{pboundaryd-1}
\end{equation}
For Region III ($x>d$), the boundary condition is

\begin{equation}
\left. \frac{1}{\epsilon_{\rm R}}\frac{\partial u}{\partial x}\right|_{x=d+0}=\left\{
\begin{array}{ll}
0\ , &\ \ \ \ \mbox{for upper PEC\ \ \ \ $z_j+\ell_j<z<z_{j+1}-\ell_{j+1}$}\\
\left.\frac{1}{\epsilon_j}\frac{\partial u}{\partial x}\right|_{x=d-0}\ ,&\ \ \ \ \mbox{for middle slot\ \ \ \ $|z-z_j|<\ell_j$}\\
0\ , &\ \ \ \ \mbox{for lower PEC\ \ \ \ $z_{j-1}+\ell_{j-1}<z<z_j-\ell_j$}
\end{array}\right.\ .
\label{pboundaryd-r}
\end{equation}
The continuity of $u$ is required along the interfaces of each slot:

\begin{equation}
\left\{
\begin{array}{ll}
\left. u\right|_{x=-d-0}=\,\left. u \right|_{x=-d+0}\ , & \ \ \ \ \mbox{for all middle slots\ \ \ \ $|z-z_j|<\ell_j$}\\
\left. u\right|_{x=d-0}=\,\left. u \right|_{x=d+0}\ , & \ \ \ \ \mbox{for all middle slots\ \ \ \ $|z-z_j|<\ell_j$}
\end{array}\right.\ ,
\label{pboundary-r}
\end{equation}
where $j=1,\,2,\,\cdots,\,N$.
Finally, the field normal derivative must be zero in Region II along the interior slot walls ($|x|\leq d$), that is,
\begin{equation}
\frac{\partial u(x,\,z=z_j\pm\ell_j)}{\partial z}=0\ ,\ \ \ \ \mbox{at all slot walls\ \ \ \ $|x|\leq d$}\ .
\label{e5}
\end{equation}
In Region I, the solution of the Helmholtz equation in Eq.\,(\ref{helm}) in this case, including the incident field, can be written as

\begin{eqnarray}
u^{({\rm I})}(x,\,z)&=&e^{ik_0(x+d)}\,e^{i\beta_0z}+e^{-ik_0(x +d)}\,e^{i\beta_0z}
\nonumber\\
&-&k_0\,\epsilon_{\rm L}\int\limits_0^{\infty}\,\frac{d\beta}{k_1(\beta)}\left[A_s(\beta)\,\cos(\beta z)+
 iA_a(\beta)\,\sin(\beta z )\right]\,e^{-ik_1(\beta)(x + d)}\ .
\label{Fourier1}
\end{eqnarray}
In Region III,
the solution of the Helmholtz equation takes the form of

\begin{equation}
u^{({\rm III})}(x,\,z)=k_0\,\epsilon_{\rm R}\int\limits_0^{\infty}\,\frac{d\beta}{k_2(\beta)}\left[B_s(\beta)\,\cos(\beta z) +
iB_a(\beta)\,\sin(\beta z)\right]\,e^{ik_2(\beta)(x - d)}\ .
\label{FourierR}
\end{equation}
In Region II, the solution of the Helmholtz equation along with the boundary condition in Eq.\,(\ref{e5}) is found to be

\begin{eqnarray}
u^{({\rm II})}(x,\,z)&=&k_0\sum_j\,\theta(\ell_j-|z-z_j|)\sum_n\,\left\{\frac{\epsilon_j}{\sigma_{sn}^j}\left[a_{sn}^j\,e^{i\sigma_{sn}^j(d+x)}-b_{sn}^j\,e^{i\sigma_{sn}^j(d-x)}\right]\,\cos[\xi_{sn}^j(z-z_j)]\right.
\nonumber\\
&+&\left.i\frac{\epsilon_j}{\sigma_{an}^j}\left[a_{an}^j\,e^{i\sigma_{an}^j(d+x)}- b_{an}^j\,e^{i\sigma_{an}^j(d-x)}\right]\,\sin[\xi_{an}^j(z-z_j)]\right\}\ ,
\label{mode1}
\end{eqnarray}
where $\xi_{sn}^j=(\pi/\ell_j)\,(n-1)$ and $\xi_{an}^j=(\pi/\ell_j)\,(n-1/2)$
are for symmetric and antisymmetric slot eigenmodes, respectively.
\medskip

By using the derivative boundary conditions in Eq.\,(\ref{pboundaryd-1}) and (\ref{pboundaryd-r}) at $x=\pm\,d$ as well as the orthogonality of the Fourier expansions in Eqs.\,(\ref{Fourier1}) and (\ref{FourierR}),
the coefficients $A_s(\beta)$, $A_a(\beta)$, $B_s(\beta)$ and $B_a(\beta)$ can be expressed as

\begin{eqnarray}
A_s(\beta)&=&\frac{1}{\pi}\int\limits_{-\infty}^{\infty} dz\,\frac{1}{\epsilon_{\rm L}}\left.\frac{\partial u^{({\rm I})}}{\partial x}\right|_{x=-d}\,\cos(\beta z)
\nonumber\\
&=&\sum_n\,\left\{\sum_j\,\frac{\ell_j}{\pi}\,\left[\left(a_{sn}^j+b_{sn}^j\,e^{2i\sigma_{sn}^jd}\right)\,Q_{sn}^j(\beta)\,\cos(\beta z_j)\right.\right.
\nonumber\\
&-&\left.\left.i\left(a_{an}^j+ b_{an}^j\,e^{2i\sigma_{an}^jd}\right)\,Q_{an}^j(\beta)\,\sin(\beta z_j)\right]\right\}\ ,
\label{pCoefAs}
\end{eqnarray}

\begin{eqnarray}
iA_a(\beta)&=&\frac{1}{\pi}\int\limits_{-\infty}^{\infty} dz\,\frac{1}{\epsilon_{\rm L}}\left.\frac{\partial u^{({\rm I})}}{\partial x}\right|_{x=-d}\,\sin(\beta z)
\nonumber\\
&=&\sum_n\,\left\{\sum_j\,\frac{\ell_j}{\pi}\,\left[\left(a_{sn}^j+b_{sn}^j\,e^{2i\sigma_{sn}^jd}\right)\,Q_{sn}^j(\beta)\,\sin(\beta z_j)\right.\right.
\nonumber\\
&+&\left.\left.i\left(a_{an}^j+b_{an}^j\,e^{2i\sigma_{an}^jd}\right)\,Q_{an}^j(\beta)\,\cos(\beta z_j)\right]\right\}\ ,
\label{pCoefAa}
\end{eqnarray}

\begin{eqnarray}
B_s(\beta)&=&\frac{1}{\pi}\int\limits_{-\infty}^{\infty} dz\,\frac{1}{\epsilon_{\rm R}}\left.\frac{\partial u^{({\rm III})}}{\partial x}\right|_{x=d}\,\cos(\beta z)
\nonumber\\
&=&\sum_n\,\left\{\sum_j\,\frac{\ell_j}{\pi}\,\left[\left(a_{sn}^j\,e^{2i\sigma_{sn}^jd}+b_{sn}^j\right)\,Q_{sn}^j(\beta)\,\cos(\beta z_j)\right.\right.
\nonumber\\
&-&\left.\left.i\left(a_{an}^j\,e^{2i\sigma_{an}^jd}+b_{an}^j\right)\,Q_{an}^j(\beta)\,\sin(\beta z_j)\right]\right\}\ ,
\label{pCoefBs}
\end{eqnarray}

\begin{eqnarray}
iB_a(\beta)&=&\frac{1}{\pi}\int\limits_{-\infty}^{\infty} dz\,\frac{1}{\epsilon_{\rm R}}\left.\frac{\partial u^{({\rm III})}}{\partial x}\right|_{x=d}\,\sin(\beta z)
\nonumber\\
&=&\sum_n\,\left\{\sum_j\,\frac{\ell_j}{\pi}\,\left[\left(a_{sn}^j\,e^{2i\sigma_{sn}^jd}+b_{sn}^j\right)\,Q_{sn}^j(\beta)\,\sin(\beta z_j)\right.\right.
\nonumber\\
&+&\left.\left.i\left(a_{an}^j\,e^{2i\sigma_{an}^jd}+b_{an}^j\right)\,Q_{an}^j(\beta)\,\cos(\beta z_j)\right]\right\}\ .
\label{pCoefBa}
\end{eqnarray}
The slot eigenmode expansion coefficients $a_{sn}^j$, $b_{sn}^j$, $a_{an}^j$ and $b_{an}^j$ in Eq.\,(\ref{mode1}) are determined from the
continuities of $u$ in Eq.\,(\ref{pboundary-r}) at $x=\pm\,d$ for individual slot.
As a result, we get for $j=1,\,2,\,\cdots,\,N$

\begin{eqnarray}
&&\,\int\limits_{z_j-\ell_j}^{z_j+\ell_j} \frac{dz}{\ell_j}\,
\left[\begin{array}{c}
\cos[\xi_{sn}^j(z-z_j)]\\
\sin[\xi_{an}^j(z-z_j)]
\end{array}\right]\,u^{{\rm I}}(x=-d,\,z)
\nonumber
\\
\nonumber\\
&=&\,\int\limits_{z_j-\ell_j}^{z_j+\ell_j} \frac{dz}{\ell_j}\,
\left[\begin{array}{c}
\cos[\xi_{sn}^j(z-z_j)]\\
\sin[\xi_{an}^j(z-z_j)]
\end{array}\right]\,
u^{{\rm II}}(x=-d,\,z)\ ,
\label{boundaryd-2}
\end{eqnarray}

\begin{eqnarray}
&&\,\int\limits_{z_j-\ell_j}^{z_j+\ell_j} \frac{dz}{\ell_j}\,
\left[\begin{array}{c}
\cos[\xi_{sn}^j(z-z_j)]\\
\sin[\xi_{an}^j(z-z_j)]
\end{array}\right]\,u^{{\rm III}}(x=d,\,z)
\nonumber
\\
\nonumber\\
&=&\,\int\limits_{z_j-\ell_j}^{z_j+\ell_j} \frac{dz}{\ell_j}\,
\left[\begin{array}{c}
\cos[\xi_{sn}^j(z-z_j)]\\
\sin[\xi_{an}^j(z-z_j)]
\end{array}\right]\,
u^{{\rm II}}(x=d,\,z)\ .
\label{boundaryd-3}
\end{eqnarray}
As for the $s-$polarization case, using the orthogonality of the slot eigenmodes, as well as
using Eqs.\,(\ref{pCoefAs})-(\ref{pCoefBa}), gives rise to the following set of $4N$ linear equations with respect to
$a_{sn}^j$, $b_{sn}^j$, $a_{an}^j$ and $b_{an}^j$

\begin{eqnarray}
&&\sum_{n^\prime,\,j^{\,\prime}}\,\left(a_{sn^\prime}^{j^{\,\prime}}+b_{sn^\prime}^{j^{\,\prime}}\,e^{2i\sigma_{sn^\prime}^{j^{\,\prime}}d}\right)\,
\left\{\frac{\epsilon_{\rm L}\ell_{j^{\,\prime}}}{\pi}\int\limits_0^{\infty} d\beta\,Q_{sn}^j(\beta)\,Q_{sn^\prime}^{j^{\,\prime}}(\beta)\,\frac{1}{k_1(\beta)}\,
\cos[\beta(z_{j^{\,\prime}}-z_j)]\right\}
\nonumber\\
&-&i\sum_{n^\prime,\,j^{\,\prime}}\,\left(a_{an^\prime}^{j^{\,\prime}}+b_{an^\prime}^{j^{\,\prime}}\,e^{2i\sigma_{an^\prime}^{j^{\,\prime}}d}\right)\,
\left\{\frac{\epsilon_{\rm L}\ell_{j^{\,\prime}}}{\pi}\int\limits_0^{\infty} d\beta\,Q_{sn}^j(\beta)\,Q_{an^\prime}^{j^{\,\prime}}(\beta)\,\frac{1}{k_1(\beta)}\,
\sin[\beta(z_{j^{\,\prime}}-z_j)]\right\}
\nonumber\\
&+&\sum_{n^\prime,\,j^{\,\prime}}\frac{\delta_{j,j^\prime}\delta_{n,n^\prime}\chi_{n^\prime}
\epsilon_{j^\prime}}{\sigma^{j^\prime}_{sn^\prime}}\,(a^{j^\prime}_{sn^\prime}-b^{j^\prime}_{sn^\prime}\,e^{2i\sigma^{j^\prime}_{sn^\prime}d})=\frac{2}{k_0}\,e^{i\beta_0z_j}\,Q^j_{sn}(\beta_0)\ ,
\label{peq5}
\end{eqnarray}

\begin{eqnarray}
&&-i\sum_{n^\prime,\,j^{\,\prime}}\,\left(a_{sn^\prime}^{j^{\,\prime}}+b_{sn^\prime}^{j^{\,\prime}}\,e^{2i\sigma_{sn^\prime}^{j^{\,\prime}}d}\right)\,
\left\{\frac{\epsilon_{\rm L}\ell_{j^{\,\prime}}}{\pi}\int\limits_0^{\infty} d\beta\,Q_{an}^j(\beta)\,Q_{sn^\prime}^{j^{\,\prime}}(\beta)\,\frac{1}{k_1(\beta)}\,
\sin[\beta(z_j-z_{j^{\,\prime}})]\right\}
\nonumber\\
&+&\sum_{n^\prime,\,j^{\,\prime}}\,\left(a_{an^\prime}^{j^{\,\prime}}+b_{an^\prime}^{j^{\,\prime}}\,e^{2i\sigma_{an^\prime}^{j^{\,\prime}}d}\right)\,
\left\{\frac{\epsilon_{\rm L}\ell_{j^{\,\prime}}}{\pi}\int\limits_0^{\infty} d\beta\,Q_{an}^j(\beta)\,Q_{an^\prime}^{j^{\,\prime}}(\beta)\,\frac{1}{k_1(\beta)}\,
\cos[\beta(z_j-z_{j^{\,\prime}})]\right\}
\nonumber\\
&+&\sum_{n^\prime,\,j^{\,\prime}}\frac{\delta_{j,j^\prime}\delta_{n,n^\prime}\epsilon_{j^\prime}}
{\sigma^{j^\prime}_{an^\prime}}\,(a^{j^\prime}_{an^\prime}-b^{j^\prime}_{an^\prime}\,e^{2i\sigma^{j^\prime}_{an^\prime}d})=\frac{2}{k_0}\,e^{i\beta_0z_j}\,Q^j_{an}(\beta_0)\ ,
\label{peq6}
\end{eqnarray}

\begin{eqnarray}
&&\sum_{n^\prime,\,j^{\,\prime}}\,\left(a_{sn^\prime}^{j^{\,\prime}}\,e^{2i\sigma_{sn^\prime}^{j^{\,\prime}}d}+b_{sn^\prime}^{j^{\,\prime}}\right)\,
\left\{\frac{\epsilon_{\rm R}\ell_{j^{\,\prime}}}{\pi}\int\limits_0^{\infty} d\beta\,Q_{sn}^j(\beta)\,Q_{sn^\prime}^{j^{\,\prime}}(\beta)\,\frac{1}{k_2(\beta)}\,
\cos[\beta(z_{j^{\,\prime}}-z_j)]\right\}
\nonumber\\
&&-i\sum_{n^\prime,\,j^{\,\prime}}\,\left(a_{an^\prime}^{j^{\,\prime}}\,e^{2i\sigma_{an^\prime}^{j^{\,\prime}}d}+b_{an^\prime}^{j^{\,\prime}}\right)\,
\left\{\frac{\epsilon_{\rm R}\ell_{j^{\,\prime}}}{\pi}\int\limits_0^{\infty} d\beta\,Q_{sn}^j(\beta)\,Q_{an^\prime}^{j^{\,\prime}}(\beta)\,\frac{1}{k_2(\beta)}\,
\sin[\beta(z_{j^{\,\prime}}-z_j)]\right\}
\nonumber\\
&-&\sum_{n^\prime,\,j^{\,\prime}}\frac{\delta_{j,j^\prime}\delta_{n,n^\prime}\chi_{n^\prime}\epsilon_{j^\prime}}
{\sigma^{j^\prime}_{sn^\prime}}\,(a^{j^\prime}_{sn^\prime}\,e^{2i\sigma^{j^\prime}_{sn^\prime}d}-b^{j^\prime}_{sn^\prime})=0\ ,
\label{peq7}
\end{eqnarray}

\begin{eqnarray}
&&-i\sum_{n^\prime,\,j^{\,\prime}}\,\left(a_{sn^\prime}^{j^{\,\prime}}\,e^{2i\sigma_{sn^\prime}^{j^{\,\prime}}d}+b_{sn^\prime}^{j^{\,\prime}}\right)\,
\left\{\frac{\epsilon_{\rm R}\ell_{j^{\,\prime}}}{\pi}\int\limits_0^{\infty} d\beta\,Q_{an}^j(\beta)\,Q_{sn^\prime}^{j^{\,\prime}}(\beta)\,\frac{1}{k_2(\beta)}\,
\sin[\beta(z_j-z_{j^{\,\prime}})]\right\}
\nonumber\\
&+&\sum_{n^\prime,\,j^{\,\prime}}\,\left(a_{an^\prime}^{j^{\,\prime}}\,e^{2i\sigma_{an^\prime}^{j^{\,\prime}}d}+b_{an^\prime}^{j^{\,\prime}}\right)\,
\left\{\frac{\epsilon_{\rm R}\ell_{j^{\,\prime}}}{\pi}\int\limits_0^{\infty} d\beta\,Q_{an}^j(\beta)\,Q_{an^\prime}^{j^{\,\prime}}(\beta)\,\frac{1}{k_2(\beta)}\,
\cos[\beta(z_j-z_{j^{\,\prime}})]\right\}
\nonumber\\
&-&\sum_{n^\prime,\,j^{\,\prime}}\frac{\delta_{j,j^\prime}\delta_{n,n^\prime}\epsilon_j}{\sigma^{j^\prime}_{an^\prime}}\,(a^{j^\prime}_{an^\prime}\,e^{2i\sigma^{j^\prime}_{an^\prime}d}-b^{j^\prime}_{an^\prime})=0\ .
\label{peq8}
\end{eqnarray}
Here, $\chi_n= 2$ for $n=1$ and $\chi_n= 1$ for $n\neq1$. Similarly,
if we limit the number of slot eigenmodes in the expansion to $M$, this constitutes a $4NM-$order inhomogeneous linear-matrix equation which can be solved exactly by matrix inversion.
Again, the symmetric and antisymmetric eigenmodes in different slots with $z_j\neq z_{j^{\,\prime}}$ are coupled to each other.

\section{Numerical Results}
\label{sec3}

To demonstrate these results,
we perform numerical calculations here for the $p-$polarization case which allows for the excitation of a surface-plasmon-polariton mode. In our numerical calculations, we have taken:
$\lambda_0=2\pi c/\omega=1.0\,\mu$m (for a large transmission), $\ell_j=\lambda_0/10=0.1\,\mu$m (deep sub-wavelength regime), $d=3.0\,\ell_j=0.3\,\mu$m, $n_{\rm L}=n_{\rm R}=n_j=1$, and the
center-to-center slot separation is assumed to be $0.98\,\mu$m. Other parameters,
such as $\theta_0$ and $N$, will be directly given in the figure captions.
\medskip

\begin{figure}[p]
\begin{center}
\includegraphics*[width=6.0in]{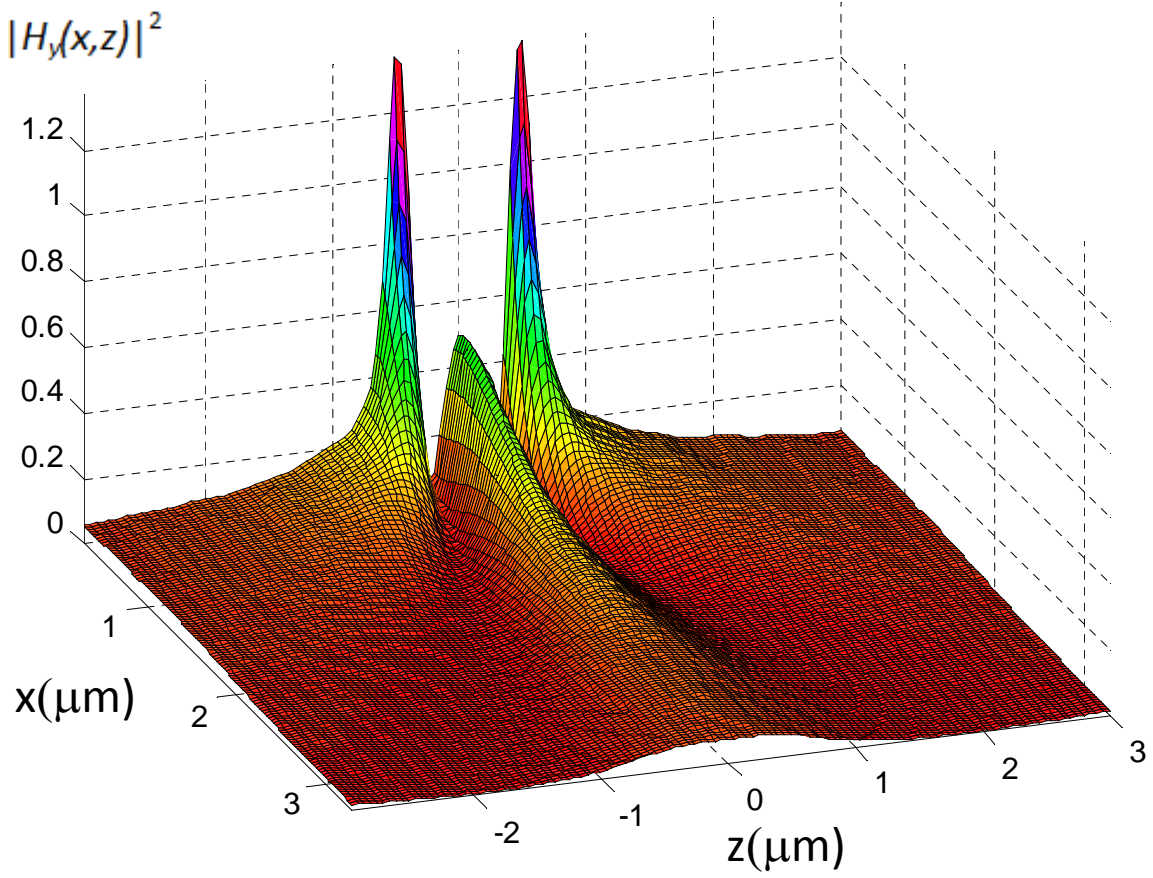}
\caption{(Color online) Three-dimensional (3D) plot of transmitted EM wave $|u(x,\,z)|^2=|H_y(x,\,z)|^2$ in Region III for $p$ polarization with $N=2$ and $\theta_0=0^{\,\rm o}$ for normal incidence.}
\label{f2}
\end{center}
\end{figure}

\begin{figure}[p]
\begin{center}
\includegraphics*[width=6.0in]{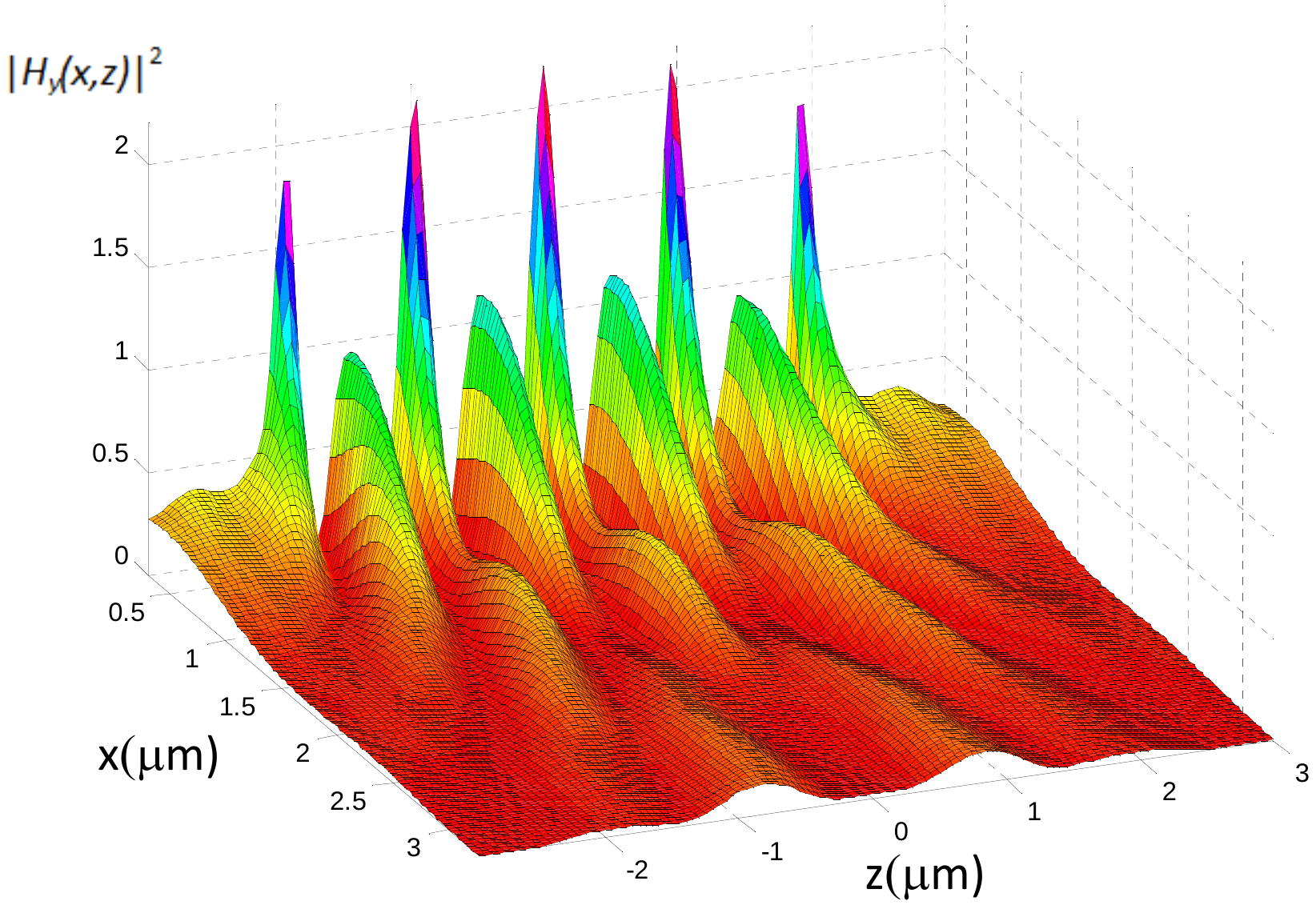}
\caption{(Color online) 3D plot of transmitted EM wave $|u(x,\,z)|^2=|H_y(x,\,z)|^2$ in Region III for $p$ polarization with $N=5$ and $\theta_0=0^{\,\rm o}$ for normal incidence.}
\label{f3}
\end{center}
\end{figure}

\begin{figure}[p]
\begin{center}
\includegraphics*[width=6.0in]{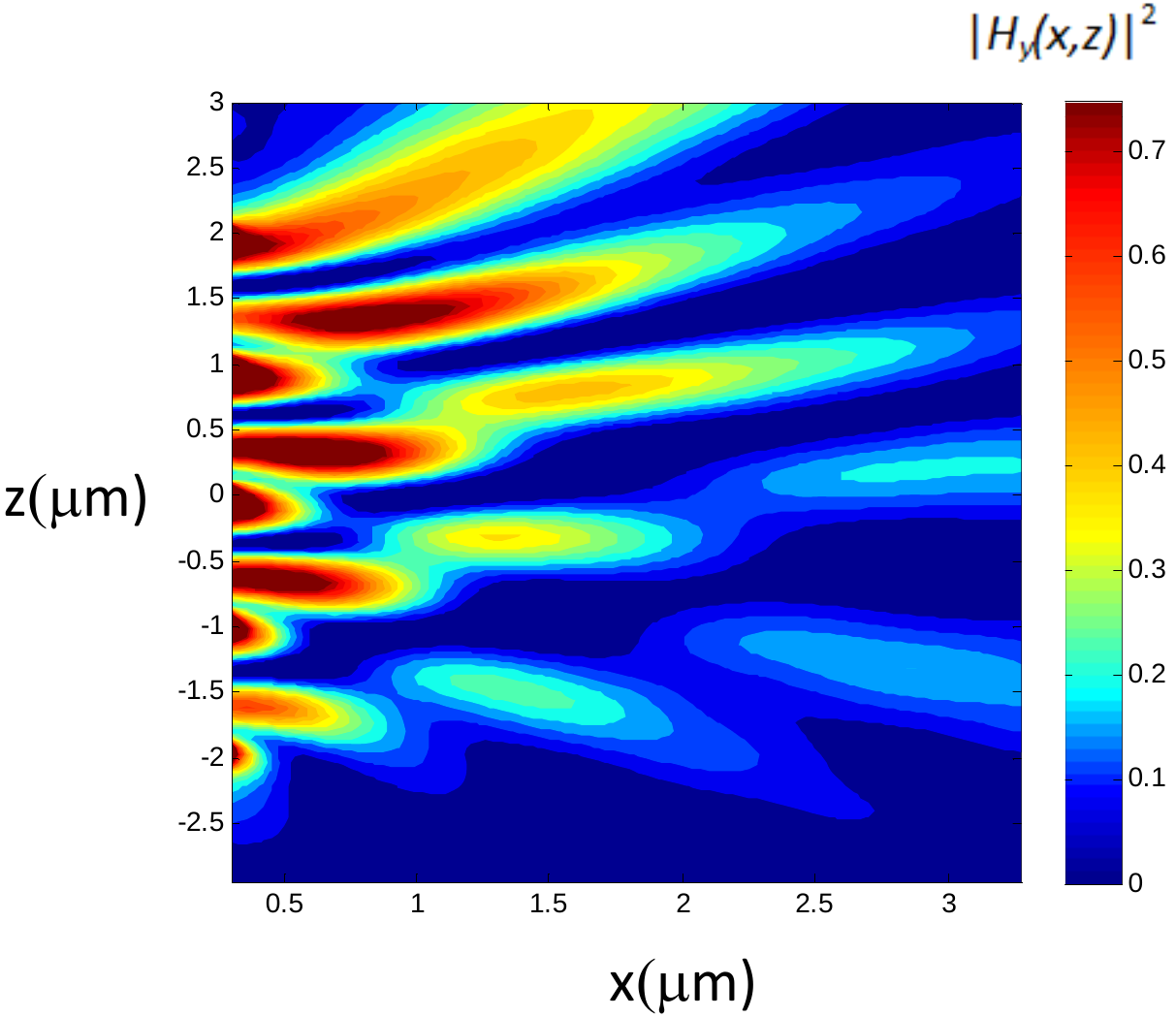}
\caption{(Color online) Two-dimensional contour plot of transmitted EM wave $|u(x,\,z)|^2=|H_y(x,\,z)|^2$ in Region III for $p$ polarization with $N=5$ and $\theta_0=60^{\,\rm o}$ for tilted incidence.}
\label{f4}
\end{center}
\end{figure}

Figure\ \ref{f2} displays the three-dimensional plot of the transmitted EM wave $|u(x,\,z)|^2=|H_y(x,\,z)|^2$ in Region III for $p$ polarization and normal incidence
with $N=2$ and $\theta_0=0^{\,\rm o}$. Here, the strong transmitted near-field close to the exits of the two slots in the $z$ direction decay very fast as it propagates away from
the slot array in the $x$ direction. At the same time, a local maximum is gradually built up in the center region between the two slots
due to strong interference of these two slot-exit waves and it becomes weaker away the slot layer in the $x$ direction.
When the number of slots is increased from two  to five in Fig.\,\ref{f3}, the strong transmitted near-fields close to the five slot exits interfere with each other strongly in the $z$ direction and
constitute a complex pattern as they propagate away from the slot array in the $x$ direction. There are four local maxima that develop eventually at the four centers between two adjacent slots as the strong slot-exit
near-fields are partially suppressed with distance away from the array in the $x$ direction. For normal incidence, although the symmetric and antisymmetric modes of different slots are coupled
to each other, the mirror symmetry with respect to the center of the system (at $z=0$) still remains. However, this mirror symmetry is broken for a nonzero incident angle, as demonstrated in Fig.\,\ref{f4} by
the two-dimensional contour plot of the transmitted EM wave $|u(x,\,z)|^2=|H_y(x,\,z)|^2$ in Region III for $p$ polarization with $\theta_0=60^{\,\rm o}$. In this case, the transmitted EM wave becomes weaker in general.
At the same time, the field strength increases with $z$ from
the lowest slot to the upper most slot, which is accompanied by an increasingly extended spatial distribution of the fields between two adjacent slots as one moves up in the $z$ direction.
\medskip

The advantage of the current approach is in casting a continuous spatial distribution problem into a small discrete coefficient matrix spanned by the slot index $j$ and the eigenmode index $n$.
In this paper, we only display the spatial distribution of a transmitted EM wave in the deep sub-wavelength regime ($\ell_j\ll\lambda_0$). However, the accuracy of our numerical calculations becomes worse
as the slot width $2\ell_j$ becomes comparable to the incident wavelength $\lambda_0$. In this case, a very large number of slot eigenmodes are required to achieve a good accuracy in a numerical calculation.

\section{Conclusions}
\label{sec4}

In conclusion, the previous analytical calculation for the diffraction of an incident plane electromagnetic wave by a single slot on a metal film has been generalized to treating
an arbitrary linear array of slots with variable slot width, slot separation and slot dielectric material.
In comparisons with the plane-wave and finite-difference time-domain methods, the calculation speed has been increased greatly by transforming the calculation of a continuous spatial distribution
of an electromagnetic field to the inversion of a small discrete coefficient matrix spanned by the slot and eigenmode indexes.
Moreover, based on a similar partial-domain method, the linearly slotted metal film with different dielectric material in each slot, which cannot be solved by using the analytical Green's function approach,
has been studied by our analytical approach. Some numerical results that demonstrate our generalized model have also been presented.

\acknowledgements

We would like to thank the Air Force Office of Scientific Research (AFOSR) for its support.

\newpage

\newpage

\newpage

\newpage

\end{document}